# Interactive Shape Perturbation

Juan C. Quiroz
Computing and Information Systems
Sunway University
Bandar Sunway, Malaysia
juanq@sunway.edu.my

Sergiu M. Dascalu
Computer Science and Engineering
University of Nevada, Reno
Reno, NV, USA
dascalus@cse.unr.edu

**Abstract** *We present a web application for the procedural generation of perturbations of 3D models. We generate the perturbations by generating vertex shaders that change the positions of vertices that make up the 3D model. The vertex shaders are created with an interactive genetic algorithm, which displays to the user the visual effect caused by each vertex shader, allows the user to select the visual effect the user likes best, and produces a new generation of vertex shaders using the user feedback as the fitness measure of the genetic algorithm. We use genetic programming to represent each vertex shader as a computer program. This paper presents details of requirements specification, software architecture, high and low-level design, and prototype user interface. We discuss the project's current status and development challenges.*

## 1 Introduction

Procedural content generation (PCG) is the algorithmic creation of game content. PCG provides the potential to reduce the cost and time to create content, while also augmenting the creativity of designers, artists, and programmers [1]. We present a PCG web application that enables users to create perturbations of 3D models. Rather than creating 3D models from scratch or from a set of polygonal primitives, we start with a well-formed 3D model and explore variations of the 3D model. The perturbations are generated with vertex shaders.

A vertex shader allows mathematical operations to be performed on the individual vertices that make up a 3D model [2]. The vertex shader performs operations on each vertex, and thus provides great flexibility to modify the position, color, texture, and lighting of individual vertices. The challenge is that making a vertex shader requires programming and computer graphics experience. In addition, even if a programmer writes a vertex shader that produces interesting results, creating a variation of that vertex shader is not straightforward.

We use an interactive genetic algorithm (IGA) to allow users to explore perturbations of 3D models. In an IGA, a user guides a search process by visually evaluating solutions and providing feedback based on personal preferences [3]. Our web application displays perturbed 3D models, the user selects the perturbation he/she likes the best, and the user feedback is used to generate new perturbations. This process is repeated until the user is satisfied. The IGA generates the vertex shaders using genetic programming (GP) [4]. In GP, computer programs are typically represented as tree structures [4], [5].

This paper makes two contributions. First, we present a web PCG application for exploring perturbations of 3D models. Our web application runs on a web browser without having to install plug-ins or any additional software. Second, we use GP to evolve vertex shaders which perturb the vertices that make up the 3D models. In this paper, we use the terms perturbations and transformations interchangeably.

The remainder of this paper is structured as follows. Section 2 describes background on genetic algorithms and related work. Section 3 lists the functional and nonfunctional software requirements for the system. Section 4 presents the use case diagram and the use cases of the system. Section 5 describes the system's architecture. Section 6 reports on the system's current status and future work.

## 2 Background

Our web application uses an existing 3D model as the seed to explore variations of the 3D model with an IGA. The user thus explores and evaluates vertex shaders by seeing the rendered result of each vertex shader applied to the 3D model and guiding the IGA with subjective feedback. Our implementation relies on genetic algorithms (GAs), interactive genetic algorithms (IGAs), and genetic programming (GP) to create the vertex shaders.

### 2.1 Genetic Algorithms

A GA is a search algorithm based on the principles of genetics and natural selection [5]. The GA maintains a population of individuals, where each individual is a potential solution to the problem being solved. In our case, each individual consists of a vertex shader program. During initialization of the GA, the individuals are created randomly. To generate a new population, parents are selected for crossover, with parent selection favoring the fittest individuals. The resulting offspring are mutated. This process repeats until a terminating condition is met.

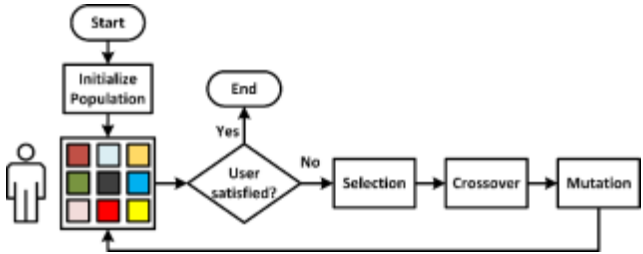

Figure 1: Flowchart of interactive genetic algorithm.

In a GA, fitness evaluation assigns a fitness value to each individual in the population. This is what drives the search process towards higher fitness solutions. When dealing with subjective criteria, such as aesthetics and user preferences, it can be difficult to create an algorithm to determine the fitness of an individual [3]. An IGA solves this by replacing the fitness evaluation with user evaluation, as illustrated in Figure 1. Thus, the IGA presents the user with individuals from the population, the user evaluates the solutions—scores, ranking, selecting the best—and the IGA proceeds with selection, crossover, and mutation.

In the canonical GA, individuals in the population are encoded using a binary string representation. In GP, computer programs are encoded and evolved by the GA [4]. The computer programs are typically represented as tree structures. In our web application we use GP to create mathematical equations to change the positions of vertices in the vertex shader.

## 2.2 Texture Evolution

Table 1 provides a summary of current literature in texture, shader, and 3D model evolution. The table indicates whether the evolutionary algorithm was driven by a user, the representation used to encode solutions, the algorithm used, and whether a target image was used to drive the evolutionary process.

Using GAs and GP for evolving images and textures in graphics was done in the early 1990s by Karl Sims [6], [7]. In the work by Sims, an IGA was used for the interactive exploration of images, textures, volume textures, 3D parametric surfaces, and animations. Sims used GP with Lisp symbolic expressions (s-expressions) as the representation. The advantage of Sims' work was that users could create a large variety of complex content without having to understand the underlying equations.

**Table 1:** Summary of texture, shader, and 3D model evolutionary systems in the current literature.

| Studies | User Interaction | Content Generated | Representation | Algorithm | Target Image |
|---|---|---|---|---|---|
| [6] | ✓ | Textures, volume textures, animations | S-expr | GP | ✗ |
| [7] | ✓ | Textures, 3D shapes, 2D dynamical systems | S-expr | GP | ✗ |
| [8], [9] | ✓ | Textures | Directed acyclic graphs | GP | ✓ |
| [10] | ✓ | Textures | Directed acyclic graphs | Steady-state GA | ✓ |
| [11], [12] | ✗ | Textures | S-expr | GP | ✓ |
| [13] | ✗ | Textures | Tree | GP | ✓ |
| [14] | ✗ | Textures | S-expr | GP | ✓ |
| [15] | ✗ | 3D textures | S-expr | GP | ✓ |
| [16] | ✗ | Textures | Tree | GP | ✗ |
| [17] | ✓ | Vertex and pixel shaders | Byte array | Linear GP | ✗ |
| [18] | ✗ | Pixel shaders | Tree | GP | ✗ |
| [19] | ✓ | Fragment shaders | Opcode numbers, numbers, and bit string | Linear GP | ✗ |
| [20], [21] | ✓ | 3D shapes | Bit string | GA | ✗ |
| [22] | ✓ | Textures | Bit string | GA | ✓ |
| [23] | ✓ | 3D Trees | Bit string | GA | ✗ |
| [24], [25] | ✓ | Vertex shaders | Bit encoded binary trees | GA | ✗ |

Another technique for automatically generating textures—without user interaction—relies on evolutionary computation to generate textures with characteristics similar to target images [8]–[11], [13], [14]. The Genshade system evolved high-level Renderman shaders to generate textures similar to a target image [8], [9]. Directed acyclic graphs were used to represent the shaders, with nodes being Renderman shader primitives. In [10], Ibrahim improved the Genshade system by incorporating a knowledge base of intermediary solutions to improve the GA search for textures that shared characteristics with target images. Ibrahim also incorporated Monte Carlo tree search to build new Renderman shaders in parallel to the GA.

In [11], the Gentropy system used GP to generated 2D textures similar to one or more target images. Gentropy used low-level texture generation with mathematical operators, noise, and turbulence effects. In [13], the evolutionary process of Gentropy was improved by including Pareto multi-objective optimization of three features: color, shape, and smoothness. In [14], textures were procedurally generated with GP and Pareto optimization of two objectives. The first objective was a mathematical model of aesthetics derived from fine art. The second objective compared the color distribution of generated images to the target image.

Hewgill and Ross used GP to evolve 3D procedural textures, where the inputs of the procedural function were the X, Y, and Z values of a vertex, and the output was an RGB value for that vertex [15]. A user selected surface points on a model and information was extracted from these surface points: coordinates, surface normal, interpolated mesh normal, and surface gradient. The GA then evaluated each procedural texture by how well its data matched the data from the surface points.

## 2.3 Vertex Shader Evolution

We refer to evolving vertex shaders as the process of using an evolutionary algorithm, such as GAs or GP, to generate vertex shaders. The outputs of this process are shaders written in a shading language. In [17], Ebner et al. used linear GP to evolve vertex and pixel shaders. Their work was limited to applying texture, color, and lighting values to pixels and vertices, without changing or displacing the positions of vertices. In [18], Howlett et al. used GP to create vertex shaders that applied coloring schemes to scenes of a 3D city environment. In [19], Meyer-Spradow et al. used GP to evolve fragment shaders that applied materials to 3D objects by using bidirectional reflection distribution functions.

In our prior work, we used an IGA to evolve equations in vertex shaders to create perturbations of 3D models [24], [25]. Our prior IGA implementation used Python-Ogre, the python binding for the OGRE graphics rendering engine. The limitations of our prior implementation were that our system (1) was a desktop application requiring users to install a large set of libraries to run the program, (2) used a bit-encoded, full binary tree representation for the individuals in the IGA. In particular, the bit-encoded, full binary tree representation allowed us to encode the equations in an array and to use a standard GA instead of using GP. However, this limited the search space of equations we could generate. The implementation described in this paper uses GP for the evolution of the vertex shaders, allowing the evolutionary process to explore a bigger space of equations.

## 2.4 Shader Editors

Due to the difficulty of shader programming, content creation tools, such as Blender or Maya, provide graphical shader development tools [26]. In these systems, shaders are created by connecting nodes and building a shade tree. Jensen et al. developed a shader editor that improved traditional shader development by automatically handling transformations between different mathematical spaces and optimizing code placement and space conversions [26].

WebGL has enabled the creation of web based shader editors. ShaderToy and GLSL Sandbox are examples of online pixel shader editors [27]–[29]. In ShaderToy, a scene consists of a fullscreen quad, with the pixel shader written using ray marching and ray casting to create graphics on the quad [28], [29]. Similarly, GLSL Sandbox is a live pixel shader editor. Both ShaderToy and GLSL Sandbox provide a platform for users to publish pixel shaders online, to allow users to edit and bootstrap off public pixel shaders. A disadvantage of ShaderToy and GLSL Sandbox is that there is no geometry in different positions over space and time in the scene. As a result, creating scenes requires implicit programming and a strong use of mathematics [28], [29].

ShaderFrog is a WebGL shader editor for vertex and fragment shaders applied to 3D objects in a 3D scene [30]. Shaders are created by writing code, by tweaking parameters of public shaders, or by using a graphical user interface to compose a shader that combines the effects of two or more shaders. ShaderFrog also provides a public repository for browsing scenes and shaders created by users.

## 2.5 Interactive Evolution of 3D Models

IGAs have also been used to evolve the parameters of algorithms that create 3D models. In [20], [21], 3D model shapes were created with the implicit surface method, which blended primitive shapes into a complex shape. The GA encoded parameters of operations applied to each primitive, such as scaling, position, rotation, tapering, shearing, twisting, etc. The GA also encoded blending parameters for combining the primitive shapes. In [23], an IGA evolved the parameters used by a procedural tree generation algorithm to create 3D trees.

In this paper we present the design and implementation of a web application for creating and exploring perturbations of 3D models. Since our system runs on a web browser, there is no need to install additional software or plugins on the client computer. We apply GP to evolve equations that are used to generate vertex shader code that changes the positions of vertices of a 3D model. In contrast, prior GP has been used to perform operations on vertex data other than position, such as colors, textures, and materials.

## 3 Requirements Specification

### 3.1 Functional Requirements

The functional requirements describe the most important behavior of the software, including user interactions and rendering processes.

1. The system shall render a 3D model.
2. The system shall divide the screen into a 3x3 grid, with a viewport for each cell in the grid.
3. The system shall use a camera for each viewport.
4. The system shall display a 3D model within each viewport.
5. The system shall allow the user to move all of the cameras simultaneously with a single set of keyboard controls and the mouse.
6. The system shall allow the user to select with the mouse the 3D model the user likes the best.
7. The system shall allow the user to step the IGA a number of generations.
8. The system shall allow the user to save a selected perturbation of a 3D model.
9. The system shall allow the user to start the IGA.
10. The system shall allow the user to load a 3D model from the file system.
11. The system shall allow the user to browse perturbations stored in a public database.
12. The system shall allow the user to use a perturbation loaded from a public database to seed the IGA.

### 3.2 Non-Functional Requirements

The non-functional requirements outline the most important constraints on the system.

1. The system shall render in real-time.
2. The system shall be implemented with HTML, JavaScript, and Three.js
3. The system shall run on a web browser with HTML5 support and hardware acceleration.
4. The server request handling shall be implemented with Python.
5. The IGA shall be implemented with the Distributed Evolutionary Algorithms in Python library.
6. The system shall be implemented using a representational state transfer (REST) architecture.
7. The system shall use GP to generate the vertex shader equations.
8. The system shall use 3D models in JSON format.
9. The system shall use time as an input to the vertex shader.
10. The system shall use addition, subtraction, multiplication, division, negation, ceiling, floor, square root, log, sine, and cosine functions in the equations.

## 4 Use Case Modeling

To gain further insight into the functionality of the system, we have divided the behavior of the web application into use cases. The use case diagram in Figure 2 outlines the controls that allow the user to interact with the system and the functionality on the server-side for running the IGA. In order to further clarify the functionality, detailed descriptions of each use case are presented.

**UC01** *Start Evolution:* The user selects the parameters of the IGA. The user then pushes the start button to display the first set of perturbations.

**UC02** *Select Model:* The user selects the perturbation the user likes best. The user can select multiple perturbations.

**UC03** *Step Generation:* The user submits the selected perturbation by clicking the Next button or by using a keyboard shortcut. This sends a request to the server to assign fitness values to the individuals in the population and create a new population.

**UC04** *Camera Control:* The user can move the camera to zoom in to the 3D model or to view the 3D model from a different angle. The camera controls will move all of the viewport cameras at the same time. The keyboard and the mouse can be used to control the camera.

**UC05** *Load Model:* The user can upload a 3D model in JSON format using a file browser dialog. After the upload completes, the 3D model will be displayed within each viewport. The user can also browse 3D models uploaded by other users, select one of the models, and load the model to the scene.

**UC06** *Save Transformation:* The user can save a transformation by first making a selection and then clicking the save button.

**UC07** *Load Transformation:* The user can load a transformation from a file or from a database of perturbations created collaboratively by users over time. Loading the transformation/perturbation will also inject the corresponding vertex shader equation to the IGA population.

**UC08** *Render Scene:* The GPU on the client device will render the scene, applying a vertex shader to each of the 3D models.

**UC09** *Initialize Population:* The server initializes the population of the IGA in response to receiving a request to start the evolutionary process.

**UC10** *Fitness Evaluation:* The selections from the user are received and used to calculate the fitness of the individuals in the IGA population. The fitness of each individual is calculated by determining the similarity between the individual and the user selections.

**UC11** *New Population*: The IGA performs selection, crossover, and mutation to generate a new population of individuals.

**UC12** *Select Subset:* Out of the large population size of the IGA, the server selects a subset of vertex shaders to return to the client for rendering on the user's screen.

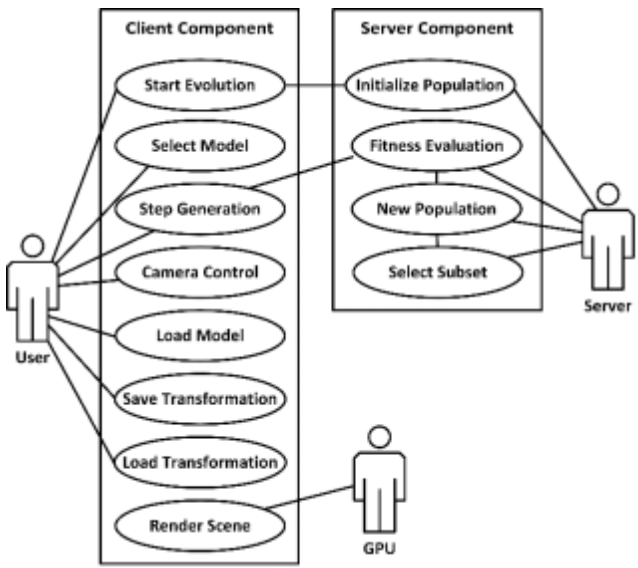

Figure 2: Use case diagram for web application.

## 5 Architectural and Detailed Design

### 5.1 Architectural Design

Figure 3 illustrates the architecture of the web application. The front-end uses WebGL for rendering graphics on the browser. WebGL is a JavaScript API for rendering 2D and 3D graphics within compatible browsers. Most importantly, no plug-ins need to be installed for the rendering to work. We use Three.js, a 3D JavaScript library, which simplifies the process of writing the WebGL code.

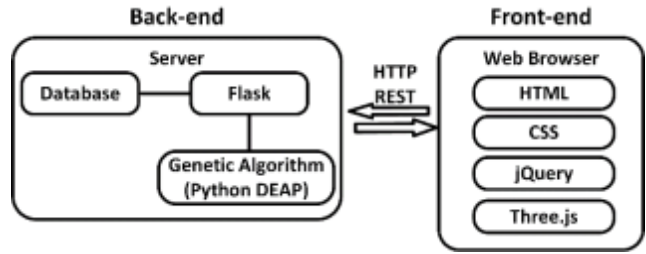

Figure 3: RESTful architecture for web application.

We used the REST architecture for the communication protocol between client and server. The RESTful web services were implemented with Flask, a web microframework for Python. Our RESTful API provides the interface for initializing the IGA population, receiving the user selections, generating a new population, saving perturbations, and loading perturbations and models. The IGA was implemented using the Distributed Evolutionary Algorithms in Python (DEAP) library. The database stores the vertex shaders saved by users and uploaded models.

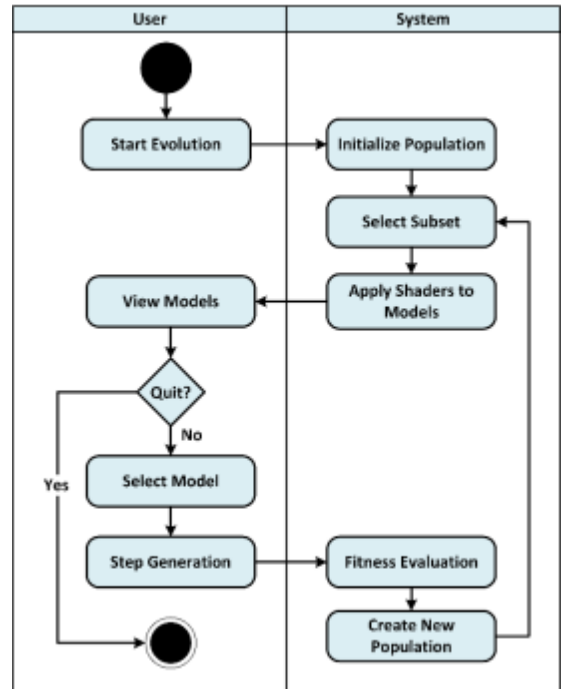

Figure 4: Activity chart for user interaction IGA.

### 5.2 System Activity Chart

Figure 4 presents an activity chart of the web application, showing the interaction between the user and the IGA. The user starts the evolution, which sends a GET request to the server to retrieve vertex shaders to apply to the models currently rendered on the user's web browser. The IGA maintains a large population size, and from this large population, a subset is selected to be evaluated by the user [25]. The user then has the option of selecting a model, and submitting this as fitness feedback to the IGA, with fitness evaluation done as in [25]. The user can repeat this process for as many generations as he/she wants.

### 5.3 Low Level Design

Figure 5 illustrates an example tree structure generated using GP. The tree structure represents the expression: $x / (x + z)$, where x and z are the x-coordinate and z-coordinate of the current vertex—we used *x* and *z* instead of *position.x* and *position.z* for brevity and readability. The internal nodes include the operators of addition, subtraction, multiplication, division, negation, ceiling, floor, square root, log, and the trigonometric functions of sine and cosine [6]. The leaf nodes include constants uniformly distributed over the half-open interval [-1, 1), a time variable, and the x, y, or z coordinate of the current vertex. The program, such as the one in Figure 5, is evaluated and it results on a scalar value, a delta. This scalar value is added to the x, y, and z coordinates of the current vertex. That is, each x, y, and z and changed by the same delta per vertex. For example, the equation from Figure 5 would be used in the vertex shader as shown in equation (1), where "*position.xyz*" represents a vector containing the x, y, and z coordinates of the current vertex.

$$position.xyz \mathrel{+}= position.x / (position.x + position.z) \quad (1)$$

Our GP uses a large population size (100), from which we select the best nine perturbations to be evaluated by the user [25]. When the user selects a perturbation as the best, the IGA interpolates the fitness of every other individual in the population based on similarity to the user selected best [25]. The fitness function calculates the fitness of an individual by taking the least square sum between the user selected best (an equation) and the individual (another equation). Since the equations have variables (x, y, z, time), we test 10,000 evenly spaced points in the interval [-10, 10] for the values of these variables. We leave the exploration of alternative fitness functions and tree similarity metrics for future work.

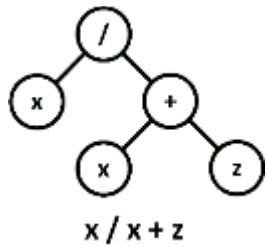

Figure 5: Sample tree structure and its evaluation.

## 6 Results

Figure 6 shows the interface of our web application. The system loads a default 3D model, consisting of a human character. Users can also upload their own 3D model, which inserts the new 3D model into the scene. A button is provided next to each model to select a model. The scene is divided into viewports, forming a 3x3 grid. Figure 6 shows an example of the interface after 8 generations. Some of the perturbations are quite destructive, such as the perturbations in the middle column of Figure 6. In some cases, the equations do not generate noticeable effects, such as the model in row two – column one.

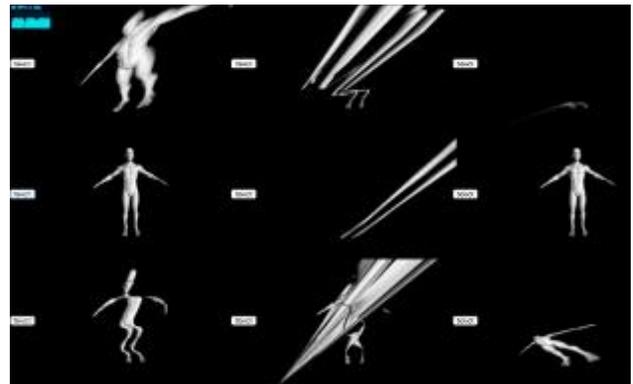

Figure 6: Main interface after evolution has started.

Figure 7 illustrates various perturbations generated with our web application. Figures 7(a)-(f) illustrate aesthetically pleasing perturbations, where the shape of the 3D model can still be appreciated. The use of trigonometric functions tends to result in curvy models, giving the models a more finished appearance. The perturbations in Figures 7(g)-(h) are quite destructive since the shape of the original model is lost and portions of the model are missing (due to vertices being translated far from the camera view space).

Equations without the time variable result in static perturbations. The disadvantage of static perturbations is that they tend to have noisy and unfinished appearances. The time variable has the potential to make perturbations interesting because of how the model geometry changes each frame. For example, the equation *sin(y * 10.0)* results in a model similar to Figure 7(f). However, Figure 7(f) is a single frame of an animation generated by the equation *sin(y * time)*. Replacing the scalar value *10.0* with the *time* variable makes the noise oscillate, which results in an interesting perturbation.

The sine and cosine functions as the root node of a GP tree result in wavy perturbations. Figure 7(e) shows the perturbation resulting from the equation *sin(y)*. Adding the time variable to this equation, *sin(y + time)*, makes the model undulate over time. Adding a multiplier to the time variable—*sin(y + 3.0 * time)*—makes the model undulate faster or slower depending on the value of the multiplier. However, making a modification such as *sin(y * time)* results in the animated noisy model from Figure 7(f).

The trigonometric equations of cosine and sine in the root node of the GP tree results in the modification to each vertex being in the range [-1, 1]. Hence, the change to each vertex is small, while still allowing for interesting results and variety. In contrast, perturbations that do not use the trigonometric functions have the potential to be destructive by exploding the model (moving vertices to positions far from the camera), see Figure 7(g)-(h). We note that a GP tree having the sine or cosine functions as child nodes in the GP tree can also result in destructive perturbations.

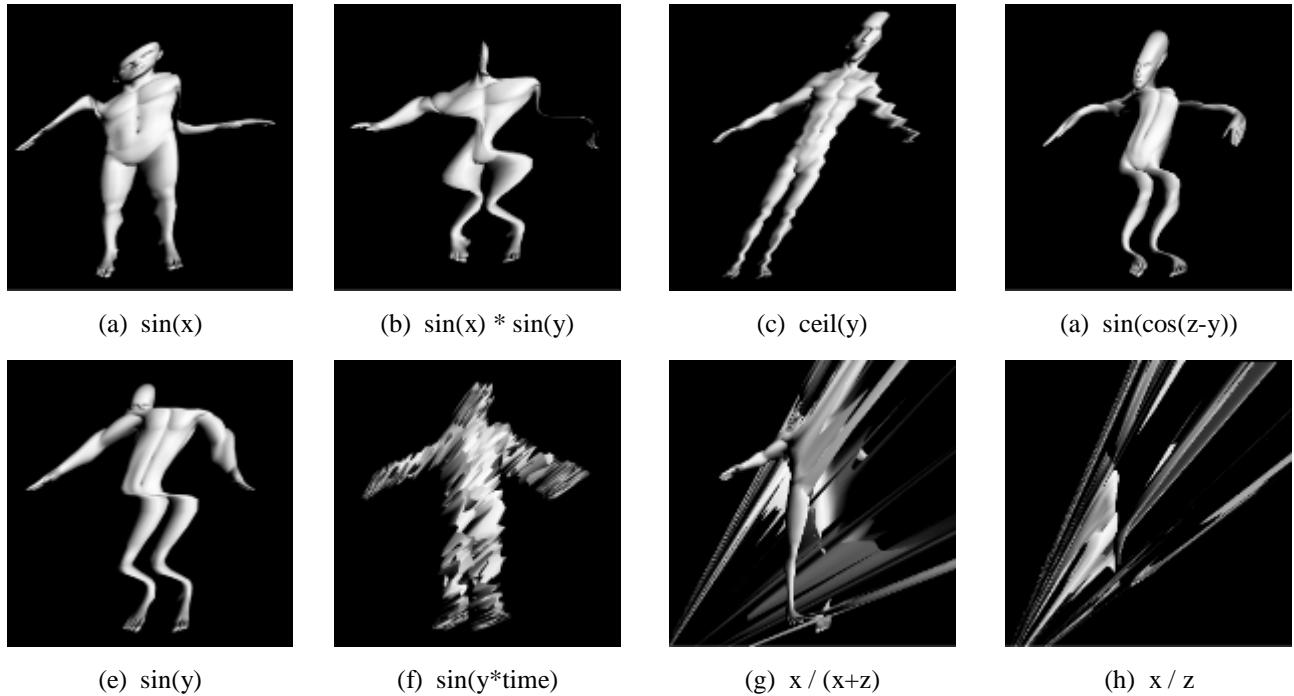

(a) sin(x)　　(b) sin(x) * sin(y)　　(c) ceil(y)　　(a) sin(cos(z-y))

(e) sin(y)　　(f) sin(y*time)　　(g) x / (x+z)　　(h) x / z

Figure 7: Example transformations generated by the corresponding equation in the vertex shader.

## 7 Conclusions and Future Work

We presented a PCG system for evolving perturbations of 3D models. The perturbations are encoded using GP, with an IGA allowing the user to quickly explore perturbations based on his/her preference. Our PCG system was implemented as a web application, allowing users to create perturbations on their web browser, without having to install libraries or plug-ins. In contrast to the online shader editors such as ShaderToy, GLSL Sandbox, and ShaderFrog, our system enables users to create vertex shaders without programming experience. All of the programming nuances are hidden from users. In fact, users do not even need to know what a vertex shader is.

A limitation of our implementation is the user interaction. IGAs tend to suffer from user fatigue, user boredom, and user inconsistency in the input provided to the GA [3]. Further work is needed to make the user interaction engaging and intuitive to users from generation to generation. A second limitation is the type of equations that can be generated with our GP implementation based on our selection of operators and terminals. The format of equation (1) also introduces a strong limitation on how the GP equations can modify the vertex data. We plan to test changing each of the x, y, and z coordinates of each vertex with different expressions. Finally, our fitness evaluation needs to be improved by using tree similarity instead of sum of least squares.